\documentclass{article}

\usepackage{arxiv}

\usepackage[utf8]{inputenc} 
\usepackage[T1]{fontenc}    
\usepackage{hyperref}       
\usepackage{url}            
\usepackage{booktabs}       
\usepackage{amsfonts}       
\usepackage{nicefrac}       
\usepackage{graphicx}
\usepackage{doi}
\usepackage{apacite}

\title{Computational Demography and Health}

\author{ \href{https://orcid.org/0000-0002-1967-123X}{\includegraphics[scale=0.06]{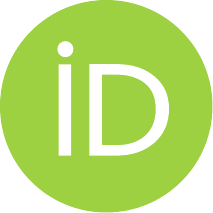}\hspace{1mm}Zack W. Almquist}\\
	Department of Sociology\\
        Department of Statistics\\
	University of Washington\\
	\texttt{zalmquist@uw.edu} \\
	\And
	\href{https://orcid.org/0000-0002-5750-5988}{\includegraphics[scale=0.06]{orcid.pdf}\hspace{1mm}Courtney Allen} \\
	Department of Sociology\\
	University of Washington\\
	\texttt{ckallen@uw.edu} \\
 	\And
	\href{https://orcid.org/0000-0003-0184-9498}{\includegraphics[scale=0.06]{orcid.pdf}\hspace{1mm}Ihsan Kahveci} \\
	Department of Sociology\\
	University of Washington\\
	\texttt{ikahveci@uw.edu} \\
}

\renewcommand{\shorttitle}{Computational Demography and Health}

\hypersetup{
pdftitle={Computational Demography and Health},
pdfsubject={cs.SI, stat.ME},
pdfauthor={Zack W.~Almquist, Courtney Allen, Ihsan Kahveci},
pdfkeywords={computational demography, computational health, big data, data science, synthetic data, social media data, online data, migration},
}

\begin{document}
\maketitle

\begin{abstract}
Recent developments in computing, data entry and generation, and analytic tools have changed the landscape of modern demography and health research. These changes have come to be known as computational demography, big data, and precision health in the field. This emerging interdisciplinary research comprises social scientists, physical scientists, engineers, data scientists, and disease experts. This work has changed how we use administrative data, conduct surveys, and allow for complex behavioral studies via big data (electronic trace data from mobile phones, apps, etc.). This chapter reviews this emerging field's new data sources, methods, and applications. 
\end{abstract}


\section*{Introduction}

Recent developments in computing, data entry, and analytic tools have changed the landscape of modern demography and health research. These changes have come to be known as computational demography, big data, and precision health in their respective fields. This emerging interdisciplinary research area comprises social scientists, physical scientists, engineers, data scientists, and disease experts. Significant developments include automated data collection such as social media or wearable health trackers; disease tracking such as Google Flu Trends or the Facebook COVID-19 Symptoms Survey; novel methods in machine learning and statistics; complex systems analysis; and so forth. This has led to prominent growth in demographic measurement, public health research, and even areas such as precision medicine (e.g., personal medicine), self-tracking, and public health initiatives. These systems have recently been used in large-scale public health outreach through COVID-19 reporting and contact tracing. This collection of methods, data, and applications is changing how we understand demography and health.

In this book Chapter, we will review major data developments, machine learning applications, statistics, and simulation models for demographic and health data and conclude with a summary and note about COVID-19. This Chapter is not a comprehensive review but a general gateway to computational demography and health --- an emerging field of study that has the potential to impact much of the social sciences.

\section*{Big data}

Big data has a rich and complicated history in the Social Sciences and has come to refer to various data sources, including but not limited to administrative data, behavioral trace data from online and offline sources, synthetic data, and geolocation data. Big data and data science are increasingly essential demographic and public health research tools. In the following section, we will review a brief background and suitability of big data and conclude with notable applications and research findings.

\subsection*{Suitability of big data for population processes and health research}

The term Big Data once described data that was too large to be stored by conventional means but has since evolved to have a more fluid definition that tries to capture the ``bigness'' of the information used by scientists across many disciplines \cite{mooney2015commentary}. This data is characterized by the \textit{three V's}: high variety, volume, and velocity. Big data are often comprised of information from multiple sources with varying original purposes (high-variety), have more cases than previously used data (high-volume), and are generated in real-time (high-velocity) \cite{mooney2015commentary, chiolero2020glossary}. The data tend to be organic, accidentally produced, for purposes other than research \cite{bohon2018demography} and reflect the increasing digitization of health and other areas of life through evolving technologies \cite{chiolero2020glossary}. Big data also describes not only the bigness of the data but data that is so large, varied, or novel that it requires new technologies and methods to understand it \cite{lazer2017data}. For example, modern computer systems like Google or Facebook can range in petabytes per day. This is in comparison to ``classic'' big data such as US Census data, tax records, and other administrative data \cite{ruggles2014big}. Some have described a \textit{fourth V} in big data: veracity. This can be considered similar to face validity and allows us to ask the fundamental question, ``Can this data answer the questions we are posing to it?"

 Big data can be noisy, with poor quality, due to its rapid production \cite{chiolero2020glossary}, and may be overwhelming to manage  \cite{raghupathi2014big, cesare2018promises}. Because big data sources are generally byproducts of digitized interactions not designed for research, they possess no defined universe and, therefore, no representativeness of any population \cite{ruggles2014big}. This creates a shallowness that has come to typify these sources \cite{bohon2018demography} as they cannot be considered neutral, objective, or complete \cite{delpierre2018big}. Many big data sources have clear population biases, such as social media platforms heavily used by women or younger populations \cite{ruths2014social}. Using data not designed for statistical inference or representative of any population presents a challenge when considering big data's use for understanding population health. 

Researchers must take steps to understand the selection bias in nonrepresentative sources that, if not calibrated, would lead to biased and invalid conclusions \cite{yildiz2017using, zagheni2015demographic}. These sources have biases that go beyond nonrepresentative population bias. For example, standard statistical approaches cannot be used due to data's network and relational aspect when using social media sources \cite{hunter2018ethical}. In addition, big data can be gleaned from platforms where individual behaviors are affected by proprietary algorithms, of which researchers have no way of correcting these influences. Furthermore, embedded researchers - those connected to such platforms - may analyze this data while correcting for algorithm influence but not be able to be transparent about their methods \cite{ruths2014social, hunter2018ethical}. Researchers encounter barriers when using big data that require advanced or novel approaches to correct bias while wielding such large data sets. These barriers are overcome with specific and advanced skills mainly missing from many graduate training programs in the social sciences and demography \cite{cesare2018promises}. This lack of training creates a slower production of robust, cutting-edge methodologies and slows the popularity and growth of big data use in demography and other social sciences. This area of research is quickly evolving, and this description is not exhaustive of all of the challenges and biases that exist when using big data. Many of these issues have existed in the social sciences, statistics, epidemiology, and other domains. Utilizing big data, collaboration, and knowledge sharing will push the field further.

Big data often includes personal information gleaned from the many digitized areas of people's everyday lives, creating ethical implications. Internet data is an increasingly used big data source with great potential to provide novel insights into demographic behaviors \cite{zagheni2015demographic}. Often referred to as trace data, this data is derived from digitized interactions from social media platforms or other digital activity and may be collected unknowingly from platform users may contain sensitive information, and their ownership remains unclear as data are sold or shared with third parties \cite{cesare2018promises}. Individuals' privacy is one element of this data that remains unclear. The right to privacy would ensure individuals' freedom from intrusion, though many digital interactions lie within an ambiguous public-private continuum \cite{hunter2018ethical}. Some data points may be collected in a manner out of the control of any researcher or individual to collect or provide consent \cite{hunter2018ethical}. Confidentiality and anonymity are also vital pieces of ethical research. While many researchers have taken steps to remove conventional identifying information, digital data are complex, and even anonymized data may expose individuals through social network ties \cite{hunter2018ethical}. Data that reflect individual actions, behaviors, or other personal attributes may not offer safeguard options to those who would not wish to contribute their data for larger research aims. This has created tensions between data protection for individuals in a community and maximizing researchers' access to big data \cite{salerno2017ethics}. 

\subsection*{Comparability of big data to other sources}

Google's Flu Trends (GFT) is the first major big data implementation for predicting health outcomes. It was developed by \cite{ginsberg2009detecting} to predict the CDC’s seasonal estimates using search queries. However, in 2013, Google stopped the service, as it infamously made the headlines after falsely predicting flu rates almost double the CDC estimates. In their work, \cite{lazer2014parable} explain the potential sources of biases embedded in the model. In addition, they show that if GFT systematically calibrated its predictions with the CDC estimates, it would result in much more accurate estimates. Following \citeauthor{lazer2017data}'s (\citeyear{lazer2017data}) terminology, we underlined the various ways of ``methodological integration'' of online data with ground truth data in this chapter.

The proponents of the Big Data Revolution often claim that as the size of data increases, the quality of information gained will also increase. However, one must remember that big data is not collected for research purposes. It contains measurement errors and selection bias \cite{zagheni2015demographic}. Recent work has questioned whether big data methods are superior to classic survey sampling for general population representativeness. \cite{bradley2021unrepresentative}. In addition, the commercial goals behind the services, such as Google’s ad services, significantly alter the data generation process over time, causing problems for replicability \cite{lazer2014parable}. Two methodological approaches are suggested for reducing the selection bias: (i) First, where ground data exist (such as administrative data), online data can be calibrated to match demographic variables. (ii) Second, where no ground data exist, difference-in-differences can reduce bias when estimating trends from online data \cite{zagheni2015demographic}. An example of this approach includes two studies using Facebook's advertising platform to predict regional public health outcomes. \cite{gittelman2015new} combines the Behavioral Risk Factor Surveillance System (BRFSS), vital statistics, and Facebook data to predict county-level health outcomes and health-related behaviors. The second, \cite{mejova2018online} employs a quasi-instrumental design to associate disease-specific marker user interests with state-level health conditions. Both results appear promising for employing online data sources for public health monitoring.  

Recent statistical developments such as Multilevel Regression with Poststratification (MRP) provide a method to mitigate selection bias in analyzing large-scale online surveys \cite{cesare2018promises, wang2015forecasting}. This method is highly related to Bayesian hierarchical models \cite{loux2019using}, which can also be used to adjust the data from a biased sample. This includes important use cases such as respondents recruited via a Facebook survey. For example, this method has been applied to the estimation of physical activity in Brazil on a large-scale cross-sectional national survey where they demonstrated that MRP consistently outperforms single-level regression models \cite{chiolero2020glossary}. 

Another approach is computational methods such as machine learning (ML) or artificial intelligence (AI) tools that have been used to reduce bias when predicting health outcomes. Modern methods of regularization (e.g., LASSO or Ridge regression) have proved useful for model selection in public health contexts \cite{young2018using}. Work focusing on real-time monitoring using social media data has leveraged ML methods. For example, \cite{joshi2020harnessing} utilized Twitter data for the early detection of thunderstorm asthma. Using a novel monitoring algorithm based on ``weighted moving average of time-between-event values.'' One of the more important uses of these procedures is that it allow for time-critical application. \emph{Time-criticality} means that the monitoring algorithm cannot wait to alert until it counts all the events in a time interval because, at that time, the outbreak will already have occurred. The frequency of events should be taken into consideration, too. 

\subsection*{Application of big data to demography and health}

\subsubsection*{Facebook}

Facebook data and access to its users have become an important resource for demographic and health data worldwide. Recently, Facebook, in conjunction with academic researchers, has been providing micro-level data (e.g., US Zip-code level data) on people's experience with COVID-19 and self-reported symptoms \cite{reinhart2021open,astley2021global}. This work has appeared in a special issue in the \emph{Proceedings of the National Academy of Sciences} and has produced dozens of peer-reviewed articles. Other important contributions include a US county-by-county social connectedness index \cite{bailey2018social} and experimental evidence on social contagions via social media posts and their effect on happiness \cite{kramer2014experimental}.

\subsubsection*{Twitter}

Twitter data, including geolocated tweets, has been used to measure place and space and provide textual content to understand opinions, trending topics, and interaction patterns. Researchers have utilized Twitter to measure migration using geolocated tweets \cite{alexander2019impact}, estimate population totals using face and age recognition on profile photos \cite{yildiz2017using}, inform on trending health topics \cite{aramburu2020Social, batbaatar2019ontologyBased}, understand housing quality \cite{chisholm2017using}, explore substance abuse \cite{stevens2020exploring}, identify food poisoning outbreaks and other acute disease events \cite{harris2017using, joshi2020harnessing}, and even examine social distancing trends during the COVID-19 pandemic \cite{younis2020social}. Other data sources have even included smart meters to understand electricity use \cite{newing2016role}, Instagram photos that give insight into codeine misuse \cite{cherian2018representations}, and Google Street View to examine the built environment \cite{keralis2020health}.

\subsubsection*{Geolocation data}

One of the most important trace data is geolocated data. It can come in many forms, including cell phones, IP location information, and mobile apps. Major marketing vendors have built large-scale data sets such as the Safegraph \cite{andersen2020early}.

Maybe the most important of these data is geolocation data from mobile phones. Mobile phone data has been used to understand the efficacy of lock-down orders during an epidemic \cite{peak2018population}, understanding mobility patterns \cite{deville2014dynamic, blumenstock2012inferring, palmer2012new}, predicting poverty and wealth \cite{blumenstock2015predicting}, and estimating de-facto versus permanent populations\cite{morrison2020estimating}. During the COVID-19 Pandemic, anonymized Google Maps data was used to understand the impact of mobility restrictions on excess deaths \cite{basellini2021linking} and hospital admissions \cite{davies2021association}. 

In demography, geolocation data has been particularly important for answering migration problems, especially short-range movements. For example, people displaced due to disaster or moving between neighborhoods.

\subsubsection*{Geolocation data and migration}

Demographers and population health researchers have long been interested in large-scale population movement and their effects on regional populations \cite{cas2014impact}. This topic has only increased in importance as climate change and rising population densities increased the frequency of natural disasters impacting human settlements \cite{bouwer2011have}. 

Researchers have used mobility data from cell phones and other electronic devices to characterize both short-term and long-term displacement after a disaster \cite{gething2011can}; however, this technique has often been limited by an inability to pair mobility data with key additional information.  In particular, many potential hypotheses could be explored if these methods also gave information about who is displaced, the household structure of displaced persons, and the nature of locations to which displaced persons migrate (e.g., identification as a home of relatives) \cite{cas2014impact}. 

Migration decisions are complex and depend on the severity of a given disaster and the demographic composition of those in the area \cite{cas2014impact}. A displacement event can last days to years, with many choosing to relocate permanently.  Prior work has compared mobility strategies among individuals living in communities that sustained different degrees of damage due to the 2004 Indian Ocean Tsunami \cite{cas2014impact}.

\subsubsection*{Natural Disasters and Displaced Populations}

There has been much work around population movement, big data, and computational methods. For example, post-disaster population movements have traditionally been very difficult to measure (e.g., early evacuation). Both voluntary and involuntary displacement can affect demographic change in a local area \cite{cas2014impact}. These demographic changes can have material consequences for affected areas  \cite{frankenberg2013education}. 
Recent work includes Facebook's Gender-Stratified Displacement Map, which aims to quantify the magnitude of displacement following disasters, describe where the displaced population has migrated, and enable the study of these population trends by gender \cite{maas2019facebook}. Facebook's Displacement Map dataset estimates how many people were displaced by a given disaster and where those people are in the period following it at an aggregate level. Specifically, the models identify a person as displaced if their typical nighttime location patterns are disrupted after the event compared to that person’s pre-disaster patterns. These patterns are obtained from a user's location history --- an optional setting on the Facebook app that users can enable that provides precise locations \cite{maas2019facebook}. The individual data is aggregated into a city-level transition matrix showing how many people are displaced from one city to another for all source cities in the disaster-affected region for each day following the event \cite{maas2019facebook}. Facebook's Data for Good program \cite{maas2019facebook} makes available displacement maps for use by humanitarian partners, and this project is a larger part of that effort.

Recent work in the ``big data" area includes following longer-term displacement and evacuation has often used administrative data, such as the Internal Revenue Service (IRS) county migration data \cite{fussell2014recovery} or, for shorter-term estimates, the Federal Reserve Bank of New York/Equifax Consumer Credit Panel (CCP) \cite{dewaard2020outmigration}. The IRS data has been used to look at the 3-4 year recovery period after Hurricane Katrina for return migration \cite{fussell2014recovery}, and the CCP data for continued estimation of displacement due to Hurricane Maria two years after \cite{dewaard2020outmigration}. Data of this nature tend to be limited either in time (e.g., the IRS data is only available yearly) or income (e.g., CCP requires the user to have a credit card).

Twitter, a popular social media application, has been used to estimate the evacuation and displacement of people following Hurricane Maria in Puerto Rico. For example, \cite{martin2020using} demonstrated that geotagged tweets provide a representative sample of individuals aged 18–54 years and complemented data collected through a survey with a sample bias toward older respondents.  Following Hurricane Maria, it was estimated that in the Twitter sample, 8.3\% of residents relocated, and nearly 4\% continued to be displaced nine months later  \cite{martin2020using}. Other work has used Twitter to characterize aspects of displacement such as destination, differential impact, and local severity \cite{yum2020mining}. 

Facebook has been used in explorations of demographic topics such as immigrant assimilation \cite{stewart2019rock}, world fertility \cite{ribeiro2020how}, and world migration stocks \cite{zagheni2017leveraging}.  Facebook's Marketing API has recently been used to estimate displacement and evacuation from Hurricane Maria in Puerto Rico  \cite{alexander2020combining}. Further, the Facebook ads system has been used to recruit survey participants for mobility research \cite{blondel2015survey}. Facebook's advertising platform has been shown to provide a good sampling frame for providing online surveys and helps with challenges of timeliness, coverage, and cost-effectiveness \cite{grow2020addressing}. Facebook users are a good cross-section of the overall population with internet access; \cite{grow2020addressing} advises stratifying by demographic characteristics, which are known to correlate with the outcome of interest and to post-stratify in the final analysis. There is growing literature on how to adjust surveys administered on the Facebook ads \cite{zagheni2017leveraging} platform or survey system \cite{feehan2019using} to estimate both online and offline populations. Finally, recent work has demonstrated that it is possible to survey evacuation behavior post-disaster via Facebook's survey infrastructure \cite{maas2020using}.

\subsubsection*{Activity-based data and health research}

Physical trace data and movements like the quantified self-movement \cite{lee2014what} have begun to provide real-time analytics on health-related activities (e.g., running or biking) as well as constant health monitoring such as heart rate or blood pressure \cite{lupton2016quantified}. Recent work using activity-based data has demonstrated that group size can affect exercise quality \cite{zeng2017let, zeng2018stay, zeng2019friending}. Other work has been done to show how you can get population estimates of health outcomes from these activity-based data \cite{almquist2019unbiased}. There is also rising interest in using self-tracking data for personal health and healthcare management \cite{ajana2017digital}. This is an exciting area of study with the potential to have big impacts on individual health and estimates of population-level health measurements (e.g., sleep studies) and our understanding of health mechanisms (e.g., time and quality of running on personal health).

\section*{Machine Learning, Statistics and Simulation}

A major challenge with big data is that much of it is not developed for research purposes but is organically created by human action patterns such as flying or using social media. Major attempts to de-bias and illicit scientifically useful information from these data sources have led to new and improved statistical and machine learning methods. 

\subsection*{Text analysis and Natural Language Processing}

Internet search terms have been analyzed to measure racism and mortality \cite{chae2015association}, estimate and predict disease trends like flu and syphilis \cite{clemente2019improved, young2018using, ginsberg2009detecting}, and understand suicide trends \cite{jimenez2020google}. 

Promising studies published based on textual analysis of social media data using NLP methods to trace health-related behaviors. The outcome of interests are quite varied, ranging from food consumption \cite{shah2020assessing}, noise pollution \cite{peplow2021noise}, vaccine-hesitancy\cite{wilson2020social}, mental health \cite{valdez2020social, zaydman2017tweeting}. There is a large body of work on using text as data in the social sciences broadly \cite{grimmer2013text}, and the potential uses in demography and health are just coming to fruition as the availability and usability of these methods mature.  

\subsection*{Database linkages}

Data linked between multiple sources (e.g., US Census data with Internal Revenue Service migration data) is a powerful resource for modern demographic and health research. Data is linked by many different methods, from hard linkages (e.g., \cite{ferrie1996new}'s work connecting 1850 to 1860 US Census data) to probabilistic linkage methods (e.g., \cite{steorts2016bayesian} has developed Bayesian methods for entity resolution, i.e., record linkages). This is an area that is changing how we think about data in the social sciences as we are able to link demographic data to major health databases (e.g., Kaiser Permanente health data records \cite{paxton2008kaiser}) or other administrative data such as the IRS migration data \cite{hauer2019irs}.

\subsection*{Public health and demography}

Public health policy researchers have studied governmental regulations and campaigns (including tweets from experts) to derive data-driven policy recommendations for public health experts.  For example, \cite{ojo2021how} compared healthcare-led Twitter hashtags with non-healthcare-led hashtags during Gun Violence and COVID-19 epidemics and showed that the former expressed more positive and action-oriented language than the latter. \cite{lohmann2018hiv} analyzed the tweets by HIV experts and their engagement and found that  fear-related language, longer messages, and including images (e.g., photos, GIFs, or videos) resulted in more shares. And, \cite{sumner2020adherence} analyzed suicide reporting news on Facebook across countries and found that closer adherence to safe-reporting practices is associated with more engagements (e.g., likes and shares).\cite{gozzi2020datadriven} conducted online surveys via Influweb, a web platform for participatory surveillance in Italy, and utilized machine learning algorithms to examine the factors driving an individual's behavior. 

\subsection*{Social Networks and Diffusion of Information and Infection}

A major growth area in computational demography is the application of empirical and simulated networks for obtaining population-level estimates of diffusion processes for public health contexts. For example, \cite{almquist2020largescale} demonstrated that it is possible to simulate homeless-to-homeless networks for understanding the diffusion process of information or infection for people experiencing homelessness in major metropolitan areas using very limited information such as the Point-in-Time survey conducted yearly on people experiencing homelessness \cite{almquist2020connecting}. Recent work by \cite{thomas2020spatial} has extended this work and others to pull apart the network mechanisms underlying COVID-19 transmission dynamics. Overall, there is extensive literature on network diffusion methods \cite{zhang2019social}.


\subsection*{Synthetic world migration data}

Historically, there has been minimal data on worldwide international migration. In the last decade, there have been two major developments in creating global migration data. \cite{abel2014quantifying} generated 5-year  migration flow estimates from sequential stock tables published by the United Nations (UN) for 1990, 2000, and 2010. More recently, \cite{azose2019estimation} estimated 5-year bilateral migration data using return and transit components with a Bayesian statistical model. Specifically, \cite{azose2019estimation} finds the ``total number of individuals migrating internationally has oscillated between 1.13 and 1.29\% of the global population per 5-year period since 1990."

\subsection*{Fertility and mortality forecasting}

Major updates to population forecasting have been developed in the last 30 years. \cite{raftery2012bayesian} has generated probabilistic forecasts for all UN countries. This has been further improved to consider uncertainty in migration counts \cite{azose2016probabilistic}. This is an important growing field of research within the computational demography community. 

\subsection*{Applications to demography and health processes}

Work by \cite{kim2020social} and others have demonstrated that social media populations and census demographics do not differ wildly. Providing evidence that data collected from social media and other online sources can provide useful evidence in a demographic and health context. Noteworthy health applications include using Twitter to monitor syphilis \cite{young2018using}, examining the effect of lock-down orders during an Ebola outbreak on mobility \cite{peak2018population}, and using Facebook data with MRP to improve behavioral health response for spatial units like the county \cite{loux2019using}. This is, obviously, a non-exhaustive list, but it provides a basic sense of this area's extensive coverage and future importance to demographic and public health research.

\section*{A Note about COVID-19}

Major innovations to computational demography and health have arisen in the last three years due to COVID-19. This includes major surveillance activities such as the partnership with Apple and Google for contact tracing \cite{michael2020covid19} and the Facebook COVID-19 Symptoms Survey \cite{reinhart2021open}. Other significant works include models and predictions of the epidemic spread \cite{jones2020transmissiondynamics} and estimates of COVID-19's impact on racial inequities \cite{wrigley-field2020us,thomas2022geographical}.

\section*{Conclusion}

Computational demography is an emerging field in demography and public health with the potential to change how we measure, model, and theorize about population and health processes. This area includes major developments in data, such as physical tracking, mobile phone data, geolocated information, specialized app development, historical text data, and classic administrative data. Statistical and computational methods are improving our analysis, prediction, and ability to harmonize this data (e.g., probabilistic record linkages). Significant work is being done to de-bias or minimally bias statistical estimates, such as the research in de-biasing and extrapolating online trace data, such as the work done in re-weighting activity-based data \cite{almquist2019unbiased}, Multilevel-Regression with Poststratification for public health data \cite{loux2019using} or demographic re-weighting for COVID-19 symptoms survey \cite{reinhart2021open}. These methods are not without controversy and detractors; see \cite{bradley2021unrepresentative} work comparing a traditional address-based sample with a large-scale Facebook sample and found that the Facebook sample with basic demographic adjustments was consistently biased for the general population. Other promising research areas are real-time response to things like natural disasters \cite{maas2020using} and health monitoring of Google Flu, Twitter trends, and Facebook surveys.

\section*{Further readings}

Readers interested in a broader review of ethical considerations in computational social science can look at the Ethics chapter in \cite{salganik2019bitbybit}. A deeper discussion on data privacy and best practices can be found in \cite{kearns2019ethical}. Further readings on using big data for health research: on privacy issues related to social media data \cite{ford2021ethical} and concerns about marginalized communities \cite{wesson2022risks}. 

Other data sources are used in demographic and health research: LinkedIn for economic migration \cite{perrotta2022openness} or gender gaps \cite{kashyap2021analysing}, bibliometric data for high-skilled return migration \cite{zhao2022return}, and Google Search Trends for contraceptive behavior \cite{berger2021covid19}. Intimate partner violence \cite{koksal2022harnessing}. A review on machine learning methods for social sciences can be found in \cite{molina2019machine}. 

Further readings on online surveys (social media, web-based, panel): an empirical assessment of different types \cite{lehdonvirta2021social}  and meta-analysis comparing response rates \cite{daikeler2020web}. Recent publications assessed the representativeness of social media samples, such as Facebook \cite{grow2022facebook} and Twitter \cite{morstatter2021sample}. The reviews on the advances on network sampling methods: respondent-driven sampling \cite{goel2010assessing} and ego-centric sampling \cite{krivitsky2022impact}.

For a literature review on computational social science \cite{edelmann2020computational} and demography \cite{mccormick2017using}. Readers interested in epistemological discussions on existing problems and broad recommendations \cite{lazer2020computational}, deriving scientific measurements from trace data \cite{lazer2021meaningful}, methodological integration across social sciences and computational science \cite{hofman2021integrating}, and mixed-methods framework for text analysis for demographic research \cite{chakrabarti2017mixedmethods}. 

For a review of recent advancements in computational approaches in migration research, see \cite{drouhot2022computational}. Crowd-sourced online genealogies present ample opportunities for demographers, for example, mortality and kinship \cite{alburez-gutierrez2022demographic}. However, one must consider the embedded bias in family lineages \cite{stelter2022representativeness}. Another rising area of research is Genome-Wide Association Studies (GWAS), thanks to large-scale open-access genomic data \cite{mills2020sociology, uffelmann2021genomewide}. Recent studies published on emerging health fields, a survey of personalized medicine literature \cite{bershteyn2022realtime}, and disparate effects of digitalized health \cite{timmermans2020technologies}.

\section*{Acknowledgments}

Partial support for this research came from a Eunice Kennedy Shriver National Institute of Child Health and Human Development research infrastructure grant, P2C HD042828, and Shanahan Endowment Fellowship, and a Eunice Kennedy Shriver National Institute of Child Health and Human Development training grant, T32 HD101442, to the Center for Studies in Demography \& Ecology at the University of Washington. Partial support for this research also came from an NSF CAREER Award \#2142964 and an ARO Award \#W911NF-19-1-0407. The content is solely the authors' responsibility and does not necessarily represent the official views of the NIH, NSF, or ARO.

\bibliographystyle{apacite}
\bibliography{com_dem_health.bib}

\end{document}